\documentclass[prl,twocolumn,showpacs]{revtex4}
\usepackage[dvips]{graphicx}
\usepackage{latexsym}
\usepackage{amsmath}

\begin{document}
\renewcommand{\ni}{{\noindent}}
\newcommand{\bit}{{\bf\cal}}
\newcommand{\dprime}{{\prime\prime}}
\newcommand{\be}{\begin{equation}}
\newcommand{\ee}{\end{equation}}
\newcommand{\bea}{\begin{eqnarray}}
\newcommand{\eea}{\end{eqnarray}}
\newcommand{\nn}{\nonumber}
\newcommand{\bk}{{\bf k}}
\newcommand{\bQ}{{\bf Q}}
\newcommand{\bN}{{\bf \nabla}}
\newcommand{\bA}{{\bf A}}
\newcommand{\bE}{{\bf E}}
\newcommand{\bj}{{\bf j}}
\newcommand{\bJ}{{\bf J}}
\newcommand{\bs}{{\bf v}_s}
\newcommand{\bn}{{\bf v}_n}
\newcommand{\bv}{{\bf v}}
\newcommand{\la}{\langle}
\newcommand{\ra}{\rangle}
\newcommand{\dg}{\dagger}
\newcommand{\br}{{\bf{r}}}
\newcommand{\brp}{{\bf{r}^\prime}}
\newcommand{\bq}{{\bf{q}}}
\newcommand{\bo}{{\bf{0}}}
\newcommand{\ha}{\hat{\bf a}}
\newcommand{\hb}{\hat{\bf b}}
\newcommand{\hd}{\hat{\bf \delta}}
\newcommand{\hc}{\hat{\bf c}}
\newcommand{\hx}{\hat{\bf x}}
\newcommand{\hy}{\hat{\bf y}}
\newcommand{\bS}{{\bf S}}
\newcommand{\bsigma}{{\vec{\sigma}}}
\newcommand{\cU}{{\cal U}}
\newcommand{\cD}{{\cal D}}
\newcommand{\bR}{{\bf R}}
\newcommand{\pll}{\parallel}
\newcommand{\sumr}{\sum_{\vr}}
\newcommand{\cP}{{\cal P}}
\newcommand{\cQ}{{\cal Q}}
\newcommand{\cS}{{\cal S}}
\newcommand{\upa}{\uparrow}
\newcommand{\dna}{\downarrow}

\title{Excitations in correlated superfluids near a continuous
transition into a supersolid}
\author{Erhai Zhao}
\affiliation{Department of Physics, University of Toronto, Toronto,
Ontario M5S-1A7, Canada}
\author{Arun Paramekanti}
\affiliation{Department of Physics, University of Toronto, Toronto,
Ontario M5S-1A7, Canada}
\begin{abstract}
\vspace{0.1cm}
We study a superfluid on a lattice close to a transition into a
supersolid phase and show that a uniform superflow in the homogeneous
superfluid can drive the roton gap
to zero. This leads
to supersolid order around the vortex core in the superfluid, with the
size of the modulated pattern around the core being
related to the bulk superfluid density and roton gap.
We also study the electronic tunneling density of states for a uniform
superconductor near a phase transition into a supersolid phase.
Implications are considered for strongly correlated superconductors.
\end{abstract}

\maketitle
The theoretical idea that $^4$He
could become a supersolid \cite{defectpapers} at low temperature
has motivated experimental studies of
defect excitations in the solid phase \cite{goodkind}, as well as
a theoretical investigation of vortices in a superfluid near a
first-order transition to a supersolid \cite{pomeau94}, as
indirect windows into the proposed supersolid phase. Recent
experimental hints \cite{chan04} for a non-classical rotational
inertia in solid $^4$He
have revived interest in this field.
The copper oxide superconductors (SC) present a different
situation, where the primary interest is in the uniform superconducting
state, but it has long been
recognized that the underdoped regime
is plagued by various competing phases
\cite{competing} involving
spin, charge, and current ordering. Such competing phases
can significantly affect the excitations in the superconducting state
if there is a continuous transition between the uniform SC and
these ordered phases. This is viewed as
one possible cause for the
anomalous excitation spectra of the underdoped cuprates.
However, continuous phase transitions
between phases with different order parameters are unusual and have
only recently begun to be explored \cite{senthil}. Further,
recent tunneling experiments \cite{hanaguri04}
in highly underdoped
Ca$_{2-x}$Na$_x$CuO$_2$Cl$_2$ ($T_c \approx 10-20$K)
suggest that the ground state in this regime
may be a supersolid rather than an insulator.
Motivated by this, we consider here
a continuous transition between a superfluid (superconductor)
and a supersolid, on a lattice,
and explore some qualitative
consequences for excitations in the uniform
superfluid (superconducting) state near such a
transition.

\begin{figure}
\includegraphics[width=3.2in]{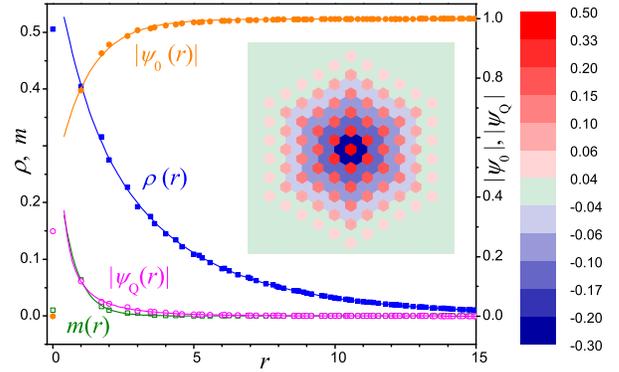}
\vspace{-4mm}
\caption{(color online) Profiles of the modulated
($\rho,|\psi_{_{\bQ}}|$) and uniform ($m,|\psi_0|$) components of
the density/superfluid order parameters around a vortex in a
superfluid near a transition to a supersolid.
Inset: Density modulations near the core.}
\label{profile}
\vspace{-4mm}
\end{figure}

\begin{figure}
\includegraphics[width=3.0in]{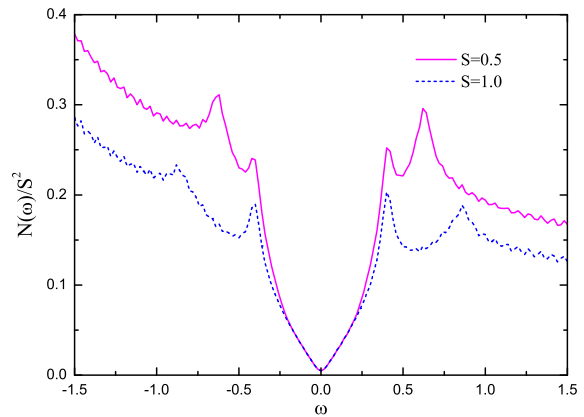}
\vspace{-4.5mm} \caption{Tunneling DOS for an $f$-wave triplet SC on
the triangular lattice. We set the spinon hopping $t_f = 1$, the
spinon pairing gap $\Delta_f = 0.4$, and set $J_\perp=J_z \approx
0.08$ in Eq.~(1) for the condensate. Low energy roton modes do not
influence the V-shaped nodal spectrum due to coherence factors. The
lower energy peaks are elastic spinon peaks, secondary peaks arise
from condensate fluctuations. At smaller $S$, the inelastic
contribution is larger while the weaker condensate dispersion shifts
the secondary peak to lower energy.} \label{stm} \vspace{-6mm}
\end{figure}

The following are our main results. (i) A continuous transition into
a commensurate supersolid phase is known to arise, with increasing
interactions, from condensation of rotons when the roton gap
vanishes. Here, we show that alternatively a uniform current flow in
the superfluid can also induce supersolid order by driving the roton
gap to zero. (ii) This current-driven collapse of the roton gap
results in a supersolid pattern emerging around the vortex core in
the uniform superfluid as shown in Fig.~\ref{profile}. The length
scale over which this modulation is significant is $R_{\rm v} \sim
D_s / E_{\rm rot}$, where $D_s$ is the bulk superfluid stiffness and
$E_{\rm rot}$ is the roton gap in the homogeneous superfluid. (iii)
We use microscopic calculations and Ginzburg-Landau (GL) theory to
obtain the long distance profiles of the modulated and uniform
components of the density and superfluid order parameter around the
vortex, as shown in Fig.~\ref{profile}. (iv) Finally, we ask whether
tunneling into a uniform SC near a transition into a supersolid phase
can probe the low energy roton excitations. Within a slave boson
approach, we show that a tunneling electron can excite the
bosonic condensate modes in the SC.
In SCs with a small superfluid density, this leads to inelastic
secondary peaks in the tunneling spectrum
as shown in Fig.~\ref{stm}.
Coherence factors however suppress
contributions from wavevectors corresponding to the roton minimum.
We conclude with possible implications for correlated SCs
such as the doped cuprate materials.

\medskip

\ni{\bf Microscopic model and Landau theory:} Consider a model of
hard-core bosons with nearest-neighbor repulsion on the triangular
lattice, \be {\mathcal{H}}\!\!= - J_\perp\sum_{\la i,j\ra}
\frac{1}{2} (b^{\dagger}_ib_j+ {\rm h.c.}) + J_z \sum_{\la
i,j\ra}(n_i-\frac{1}{2})(n_j-\frac{1}{2}). \label{model} \ee For
large $J_\perp/J_z\equiv \Delta$ the ground state of this
Hamiltonian is a uniform superfluid. In the interaction dominated
regime, for $\Delta\lesssim 0.2$, quantum Monte Carlo simulations
\cite{sspapers} have shown that the uniform superfluid is unstable
to density-wave modulations at wavevectors $\pm \mathbf{Q} \equiv
(\pm 4\pi/3,0)$, leading to a supersolid. Through the standard
mapping ($S_i^+\!\! =\!\! b_i^\dagger, S_i^-\!\! =\!\! b_i,
S_i^z\!\! = \!\!n_i\!\!-\!\!1/2$), this Hamiltonian is equivalent to
an $S=1/2$ XXZ spin model with ferromagnetic in-plane interaction
$J_\perp$ and an antiferromagnetic out-of-plane interaction $J_z$.
Spin-wave calculations on this XXZ model \cite{murthy97}, valid at
large-$S$, show that the phase transition into the supersolid is
caused by the vanishing of the roton gap at $\pm\bQ$, which leads to
roton condensation. They also capture the precise structure of the
supersolid phase of model (1) which is now known from direct
numerics for $S=1/2$ \cite{sspapers}. Further, both techniques agree
that the transition from the superfluid to the supersolid is {\it
continuous} which is also indicated from a GL analysis outlined in
Melko {\it et al.} \cite{sspapers}. Since we are interested in the
consequences of a continuous superfluid to supersolid transition, we
focus on this specific model. We will use (semi)classical analyses
of the XXZ model, supplemented by GL theory, as a reliable guide to
the physics of model (1).

Keeping the relevant wavevectors $\bq=\bo,\pm\bQ$ to describe the
superfluid to supersolid transition, the superfluid order,
$\psi(\br)$, and the deviation of the density from half-filling,
$\delta n(\br) = n(\br)-1/2$, can be expressed as $\psi(\br) \sim
\psi_0(\br) + \psi_{_\bQ}(\br) e^{i\bQ\cdot\br} +
\psi_{_{-\bQ}}(\br) e^{-i\bQ\cdot\br}$ and $\delta n(r)  \sim m(\br)
+ \rho(\br) e^{i\bQ\cdot\br} + \rho^*(\br) e^{-i\bQ\cdot\br}$. The
complex numbers $\psi_0, \psi_{_{\pm\bQ}},\rho$ and the real number
$m$ are, thus, order parameters in the GL theory
\cite{sspapers,thankashvin,footnote.sz}. For the model (1), $\psi_0$ is nonzero in, both,
the superfluid and the supersolid, while $\psi_{_{\pm\bQ}},m,\rho$
are nonzero only in the supersolid.
Since we are interested in a
hard-spin formulation here, it is convenient to work with the phase
and amplitude of the superfluid order parameters $\psi_0,
\psi_{_{\pm\bQ}}$. Terms such as $\psi_{_{\bQ}}^2
\psi^*_{_{-\bQ}} \psi^*_0$ and $\psi^*_{_{\bQ}}\psi^*_{_{-\bQ}}
\psi^2_0$ in the GL functional lock the phases of these different components,
so that $\psi_0 =|\psi_0| e^{i\varphi}, \psi_{_{\pm\bQ}} =
|\psi_{_{\pm\bQ}}| e^{i\varphi}$ are determined by three amplitudes
and a single phase variable $\varphi$. The amplitudes are
constrained by the $|\psi_0|^2 + |\psi_{_\bQ}|^2 +
|\psi_{_{-\bQ}}|^2 + m^2 + 2 |\rho|^2 =
1$, which fixes the spin length (to unity).

The GL functional $f=f_{\rho}+f_{m}+f_{\psi} + f_c$, where
\bea
f_{\rho}&=&\alpha_\rho |\rho|^2\! +\! g_\rho
|\nabla\rho|^2\! +
\! u_\rho |\rho|^4 \! +\!  w_\rho  \mathrm{Re}(\rho^6) \ldots \label{f2} \\
f_{m}&=&
\alpha_m m^2\! +\! g_{m} (\nabla m)^2
\!+\! u_m m^4 \ldots \label{f3}\\
f_{\psi}&=&\varrho_s |\nabla\varphi|^2+ \ldots \label{f4} \\
f_c&=& - \lambda_1 \varrho_s^2 |\rho|^2 |\nabla \varphi|^2 +
\lambda_2 m\mathrm{Re}(\rho^3) + \lambda_3 [\rho^2 (\psi_{_{\bQ}}
\psi^*_0 + \nn \\
&\,& \psi^*_{_{-\bQ}}\psi_0) + \mathrm{c.c.}] + \lambda_4 [m \rho \psi_{_{-\bQ}} \psi^*_0 +
\mathrm{c.c.}] \ldots \label{f5}
\eea
where we have only displayed terms relevant to our analysis below.
These terms respect the particle-hole and lattice symmetries of the
Hamiltonian in Eq.~(1), and could be, in principle, derived using
functional integral methods as outlined for a related model in
Ref.\cite{frey}. Here we treat the
coefficients
as phenomenological parameters. In Eq. (\ref{f2}),
$\alpha_\rho \equiv a(\Delta-\Delta_{c0})$ measures the distance to the
superfluid-supersolid transition point with $a
>0$. Spin-wave theory and Monte Carlo calculations indicate $w_\rho
< 0$, which prefers $\rho$ to be real in the supersolid. In $f_c$,
the term $\lambda_1$ couples the modulated
solid order and the superflow current $\varrho_s \nabla \varphi$.
The $\lambda_{2,3,4}$ terms induce effective ``magnetic fields'' upon the
different order parameters.

\medskip

\ni {\bf Current driven collapse of the roton gap:}
For superfluids close to a supersolid instability, external
perturbations that suppresses the kinetic energy relative to the
interaction energy can drive the system into the supersolid phase.
Consider a uniform superflow, say in the $x$ direction,
which introduces a phase difference between neighboring sites,
$\delta=\phi_j-\phi_i$. The phase gradient effectively suppresses
the nearest-neighbor kinetic energy from $\Delta$ to
$\Delta\cos(\phi_i-\phi_j)$. Carrying
through the calculation of spin-wave fluctuations around such
a state with superflow, we find a spin-wave dispersion
with a roton gap $
\omega(\mathbf{Q})=1.5\sqrt{6\tilde{\Delta}(\tilde{\Delta}-1/2)}$, where
$\tilde{\Delta}\equiv\Delta[2\cos(\delta/2)+\cos\delta]/3$, reduced by the
current flow. The point where the roton gap collapses is thus
shifted upward to $\Delta_c\simeq
\Delta_{c0}[1+(\delta/2)^2]$, where in the last step we assumed
$\delta\ll 1$. Recalling the current density $J \sim \sin\delta
\approx \delta$, we conclude the boost of $\Delta_c$ is
quadratic in the supercurrent density close to the transition,
and the superflow induces supersolid ordering.

In the GL approach, superflow-induced supersolid ordering can be
described phenomenologically by the coupling term $\lambda_1$
between the
supercurrent and the supersolid order parameter $\rho$ in
Eq.(\ref{f5}).
This term effectively shifts $\alpha_\rho$ to
$\alpha_\rho'=\alpha_\rho -\lambda_1 J_s^2$, where the supercurrent
$J_s = \varrho_s |\nabla \varphi|$. We thus see that one
can induce a transition from a superfluid ($\alpha_\rho' > 0$) into
a supersolid ($\alpha_\rho' < 0$) by a superflow which leads
to a sufficiently large $|\nabla \varphi|$. The
``critical current density'' required to obtain the supersolid
is $J_c=\sqrt{\alpha_\rho/\lambda_1}$.
This is analogous to superflow
induced spin-density modulation \cite{subirdemler} and
gradient couplings inducing competing phases \cite{balatsky}
considered earlier
only within GL theory. We next
turn to implications of this for a vortex in the superfluid phase.

\medskip

\ni {\bf Vortex in the superfluid phase:} For a superfluid vortex,
the current density increases upon approaching the core as $J_s =
\varrho_s/r$. The above result then indicates that the superflow
would cause a supersolid pattern to be stabilized in a region with
characteristic radius $R_v= \varrho_s \sqrt{\lambda_1/\alpha_\rho}$
around the core. It is clear from the GL functional that
$\alpha_\rho \sim E_{\rm rot}^2$, since the roton gap determines the
distance to the transition. We thus identify $R_v \sim
\varrho_s/E_{\rm rot}$. We have confirmed this by microscopic
calculations where we have evaluated the roton gap, the superfluid
stiffness $D_s$ (which has energy units and can be viewed as
$\varrho_s/m^*$ with an effective mass $m^*$) and the ``critical
current'' using spin-wave theory.

To study, further, the interplay of the various order parameters
near a vortex, we numerically studied model (\ref{model}) in the
presence of an orbital magnetic field, by replacing $t$ in Eq.
\eqref{model} with $t_{ij}=t\exp(i\int_j^i\mathbf{A}\cdot
d\mathbf{l})$. We considered an $L\times L$ lattice with periodic
boundary conditions, and a vector potential corresponding to a
uniform magnetic field with the total flux chosen to be one flux
quantum. The ground state then
corresponds to a single vortex within the cell. Transforming
to an effective spin model, and treating
the spins as classical unit vectors, we found the ground
state spin configuration $\{\mathbf{S}_\br\}$ using a simulated
annealing algorithm. The ground state corresponds to a single vortex
with its core on a site of the lattice. The local order parameters
at each $\br$, i.e. $\rho$, $m$, $\psi_0$, and $\psi_{_{\pm\bQ}}$,
were extracted from $\{\mathbf{S}_{\br}\}$ by coarse graining over
six nearest neighbor sites.

The inset to Fig.~\ref{profile} shows the map of the boson density
$\delta n(\br)$ around a vortex for $\Delta=0.505$ --- it is clear
from this that supersolid order is induced in the vicinity of the
vortex core as expected from earlier arguments. Each data point in
the plotted profiles in Fig.~\ref{profile} represents the angular
average taken at a fixed distance from the core. It is apparent
that as the superflow induces a nonzero $\rho$, the
order parameter $|\psi_0|$ is
suppressed. At the same time, $|\psi_{_{\pm\bQ}}|$ and $m$ are
generated but they are insignificant except very near the vortex core.
We explain
these features below using GL theory.

Outside of a few lattice spacings from the vortex center $m\simeq 0,
\psi_{_{\pm\bQ}} \simeq 0$ and we will ignore them to begin with.
Assuming $\rho$ is small, and $|\psi|$ is almost constant, the free
energy density relevant to $\rho$ reduces to $ f_{\mathrm{far}}
\approx (\alpha_\rho -\lambda_1 \varrho_s^2
|\nabla\varphi|^2)|\rho|^2 + g_\rho |\nabla\rho|^2 +u_\rho |\rho|^4+
\ldots$, where we have dropped the higher order $w_{\rho}$ term but
implicitly use $w_{\rho} < 0$ to consider only real $\rho$ solutions
below (the numerical solution for the vortex yields an almost purely
real $\rho$). It is clear that for small $r$,
$|\nabla\varphi|
\sim 1/r$ is large and renders the state with $\rho=0$ unstable. Minimizing
$f$ with respect to $\rho^*$, introducing $s(r)=\rho(r)2u_\rho
/\alpha_\rho $, one finds $s(r)$ obeys the differential equation
$\xi_{\rho}^2 [d^2s/dr^2+(1/r)(ds/dr)] - s^3 +(R_v^2/r^2-1) s=0$.
Here $\xi_{\rho}=\sqrt{g_{\rho}/\alpha_\rho}$ is the coherence
length for $\rho$, and $R_v= \varrho_s \sqrt{\lambda_1/\alpha_\rho}$
as defined before is the length scale associated with the supersolid
order. The solution to this equation at large $r$ is $s(r)\sim
e^{-r/\xi_{\rho}}/\sqrt{r}$. This solution breaks down close to the
vortex core as $\rho$ saturates to a finite value at $r=0$. The
numerical result for $\rho(r)$ can be fit well over the region $r >
1$ with $\rho(r)=Ae^{-r/\xi_1}\tanh(\sqrt{\xi_2/r})$, as shown by
the solid line in Fig.~\ref{profile}. The length scale $R_v \sim
[\la r^2 \rho(r) \ra / \la \rho(r) \ra]^{1/2}$, where the $\la.\ra$
denotes the spatial average with the vortex at the origin. Once we
have a nonzero $|\rho|^2$ (far from the core) it suppresses, via the
hard-spin constraint, the uniform superfluid order
parameter $|\psi_0|^2$ with $\left[ |\psi_0|^2 + 2 |\rho|^2 \right]
\approx 1 $ since $m,\psi_{_{\pm\bQ}} \simeq 0$. Next
we turn to these order parameters which are small everywhere.
We see from Eq.~(\ref{f5}) that $\lambda_2 \mathrm{Re}(\rho^3)$
acts as a magnetic field for $m$, while
$\lambda_3 \rho^2 \psi^*_0$ induces
$\psi_{_{\bQ}}$. Thus we expect $m \sim \mathrm{Re}(\rho^3)$ and
$|\psi_{_{\pm \bQ}}| \sim |\rho|^2 |\psi_0|$ (except very close
to the core), as borne out by our numerics.

\medskip

\ni {\bf Tunneling:}
We next analyze electron tunneling
into the uniform SC to see if it probes the low-energy
roton excitation when the system is on the brink of becoming
a supersolid. We use a slave boson formulation, writing the electron
operator
as $c_\sigma(\br,\tau) = f_\sigma(\br,\tau) b(\br,\tau)$ where
$f_\sigma$ is a spinful fermion (spinon) and $b$ is a charged boson.
Assuming that gauge fluctuations and interactions are innocuous
in the superconducting state, the electron Green function can be factorized
as $G^c_\sigma(\br,\tau)= G^f_\sigma(\br,\tau) G^b(\br,\tau)$.
In mean-field theory, when the boson is condensed, we can
replace $G^b(\br,\tau) \sim \la
b \ra^2$, so that $G^c \sim G^f$,
their ratio being the condensate density $\la
b\ra^2$. The electron tunneling spectrum thus reflects the tunneling
density of states (TDOS) associated with the spinon spectrum.

Going beyond mean field theory, we
assume that the bosons are described by an interacting Hamiltonian
such as (1) and evaluate the boson Green
function using spin-wave theory. Combining this with the
spinon Green function, for a model with spinon pairing,
leads to the TDOS
\bea
&& \!\!\!\!N_c(\omega) = \frac{S^2}{V} \sum_\bk \left( v^2_\bk \delta(\omega
+ E_\bk) + u^2_\bk \delta(\omega - E_\bk) \right)
\label{tdos} \\
\!\!\!\!\!\!&& \!\!\!\!\!\!\!+ \frac{S}{2 V^2} \sum_{\bq\neq 0,\bk} \gamma^2_\bq
\left[ v^2_\bk \delta(\omega
+ E_\bk + \Omega_\bq)
+ u^2_\bk \delta(\omega - E_\bk - \Omega_\bq) \right]. \nn
\eea
Here $u_\bk^2,v_\bk^2$ are the usual superconducting coherence factors
associated with spinon pairing, while
$\gamma^2_\bq$ is the coherence factor associated
with the Bogoliubov transformation for diagonalizing the boson model
\cite{footnote.bog}.
The first term, which dominates at large-$S$, reflects the mean field
result where the condensate does not fluctuate.
The second term
represents inelastic processes where tunneling
inserts a spinon and excites condensate
fluctuations.

To show the effects of condensate fluctuations in a {\it nodal} SC
on the triangular lattice, we plot in Fig.~\ref{stm} the TDOS
$N_c(\omega)$ obtained for $f$-wave triplet pairing of spinons
\cite{footnote.details}. The two main features of the spectrum in
Fig.~(\ref{stm}) are: (i) At low energy the nodal spectrum is nearly
unaffected by the condensate fluctuations. (ii) At higher energies,
condensate fluctuations lead to a broad secondary peak. This arises
from $\bq$ regions where there is a large density of states for
condensate excitations. These remain true even in the presence of
low-energy rotons since the $\gamma^2_\bq$ coherence factor tends to
reduce the contribution to the TDOS coming from near the roton
wavevector $\bQ$. The energy difference between ``elastic'' and
``inelastic'' peaks scales with the superfluid stiffness.

\medskip

\ni {\bf Experimental implications:}
Our results for the superflow induced supersolid, the
vortex core size scaling and the existence of secondary tunneling
peaks rely mainly on the existence of a continuous superfluid-supersolid
transition and a small superfluid density, largely independent
of the detailed microscopics. Here we
consider some implications for the cuprate superconductors.
Tunneling measurements on near-optimal superconducting
Bi$_2$Sr$_2$CaCu$_2$O$_{8+\delta}$ ($T_c = 89$K) in a
$B=5$T magnetic field see a modulated TDOS around the vortex
core \cite{hoffman}. Although different explanations have been proposed
for this pattern (e.g., Refs.\cite{bartosch,tesanovic,kivelson}),
it is possible that over the large region (with radius $\sim 75$\AA)
in which the modulated pattern appears
the system may be better viewed as a supersolid
(which appears to be seen at low doping and $B=0$ in
a related cuprate \cite{hanaguri04}).
If this supersolid has a charge modulation,
the roton in this system at $B=0$ could be observed using
inelastic X-ray scattering. We expect
that the roton energy should scale as $E_{\rm rot}
\sim D_s/R_v$
where $D_s$ is the bulk superfluid stiffness
and $R_v$ is the radius of the region around the
vortex core exhibiting a checkerboard modulation. Further,
the observed modulation region around the vortex core
should grow with underdoping upon approaching the supersolid.
Both predictions could be tested experimentally.
We have also shown
that condensate fluctuations can lead
to inelastic secondary peaks at higher energies, while leaving
the low energy V-shaped spectrum unaffected. This is broadly
consistent with tunneling data in
the cuprates where regions with a small superfluid
density and broad secondary tunneling peaks (so-called
``large gap'' regions)
coexist with ``small gap'' regions of larger superfluid density
and single peaks \cite{fang}. However, the fact that the
high energy structure carries information about the ordering wavevector
in the cuprates \cite{mcelroy05} does not appear to be a feature of
the model we have
studied. This may need coupling to disorder which can
simultaneously lead to low superfluid density and supersolid order in
certain regions and also to scattering of quasiparticles off the
disorder induced supersolid.
Further, one needs to address
possible bond-order instabilities \cite{bartosch,podolsky,vafek}.
We hope our work will also stimulate experiments to study supersolids
and vortices \cite{congjun} using cold atoms
on optical lattices, and vortices in superfluid
$^4$He near freezing pressure.

We thank J.C. Davis, T. Hanaguri, H. Hoffman, H.-Y. Kee, Y.-B. Kim,
Y.-J. Kim, S. Sachdev, A. Vishwanath, and Z. Wang, for useful
discussions and correspondence. This work was supported by a startup
grant and a Connaught grant from the University of Toronto.


\begin{thebibliography}{999}

\bibitem{defectpapers}
A.F. Andreev and I.M. Lifshitz, Sov. Phys. JETP {\bf 29}, 1107 (1969);
G. Chester, \pra {\bf 2}, 256 (1970); A.J. Leggett, \prl {\bf 25}, 1543
(1970).

\bibitem{goodkind}
G. Lengua and J.M. Goodkind, J. Low Temp. Phys.
{\bf 79}, 251 (1990).

\bibitem{pomeau94}
Y. Pomeau and S. Rica, \prl {\bf 72}, 2426 (1994).

\bibitem{chan04}
E. Kim and M.H.W. Chan, Nature {\bf 427}, 225 (2004);
E. Kim and M.H.W. Chan, Science {\bf 305}, 1941 (2004).

\bibitem{competing}
For representative
reviews, see
S. Kivelson {\it et al.},
\rmp {\bf 75}, 1201 (2003); E. Demler {\it et al.},
{\it ibid} {\bf 76}, 909 (2004);
L. Balents {\it et al.}, cond-mat/0504692;
M. Franz {\it et al.}, \prb {\bf 66}, 054535 (2002);
A. Melikyan and Z. Tesanovic, {\it ibid} {\bf 71}, 214511 (2005).

\bibitem{senthil}
T. Senthil {\it et al.}, Science {\bf 303}, 1490 (2004); L. Balents
{\it et al.}, \prb {\bf 71}, 144508 (2005).

\bibitem{hanaguri04}
T. Hanaguri {\it et al.}, Nature {\bf 430}, 1001 (2004).

\bibitem{sspapers}
D. Heidarian and K. Damle, \prl {\bf 95}, 127206 (2005); R. G.
Melko {\it et al.}, {\it ibid} {\bf 95}, 127207 (2005);
S. Wessel and M. Troyer, {\it ibid} {\bf 95}, 127205 (2005);
M. Boninsegni and N. Prokofiev, {\it ibid} {\bf 95}, 237204 (2005).

\bibitem{murthy97}
G. Murthy {\it et al.}, \prb {\bf 55}, 3104 (1997).
Spin wave theory finds a supersolid phase for $\Delta <
\Delta_{c0} = 0.5$.

\bibitem{thankashvin}
We are indebted to Ashvin Vishwanath for pointing out
that $\psi_{_{\pm\bQ}}$ is nonzero in the supersolid and would
be generated by the couplings $\lambda_{3,4}$ in the
GL theory.

\bibitem{footnote.sz}
In the XXZ language, these
order parameters correspond to the spin components
$S^z,S^+,S^-$ at $\bq=\bo,\pm\bQ$.

\bibitem{frey}
E. Frey and L. Balents, \prb {\bf 55}, 1050 (1997).

\bibitem{subirdemler}
E. Demler {\it et al.}, \prl {\bf 87}, 067202 (2001).

\bibitem{balatsky}
Z. Nussinov {\it et al.}, cond-mat/0409474.

\bibitem{footnote.bog}
The coherence factor $\gamma^2_\bq=(\cosh 2\theta_\bq + \sinh 2\theta_\bq)$
with $\cosh 2\theta_\bq = \epsilon_\bq/\Omega_\bq$ and $\sinh 2\theta_\bq
= \Delta_\bq/\Omega_\bq$, where $2 \epsilon_\bq=
\sum_{\hd} [J_\perp (2-\cos \bq\cdot\hd) + J_z \cos \bq\cdot\hd]$,
$2 \Delta_{\bq} = \sum_{\hd} [(J_\perp + J_z) \cos \bq\cdot\hd]$, and
$\Omega^2_\bq = \epsilon^2_\bq - \Delta^2_\bq$. Here $\hd$ are
unit vectors joining nearest neighbors on the triangular lattice.

\bibitem{footnote.details}
Details of the spinon Hamiltonian are unimportant.
Since Coulomb interactions could modify the
condensate spectrum at small $q$, we impose
$|\bq| > q_c \sim 1$ in Eq.~(\ref{tdos}). We have checked
that including the plasmon mode (dispersing as $\sim \sqrt{q}$)
and summing over all $\bq$ in Eq.~(\ref{tdos})
does not affect our results qualitatively.

\bibitem{hoffman}
J.E. Hoffman {\it et al.}, Science {\bf 295}, 466 (2002).

\bibitem{fang}
A.C. Fang {\it et al.}, \prl {\bf 96}, 017007 (2006).

\bibitem{mcelroy05}
K. McElroy {\it et al.}, \prl {\bf 94}, 197005 (2005).

\bibitem{bartosch}
L. Bartosch {\it et al.},  cond-mat/0502002.

\bibitem{tesanovic}
M. Franz {\it et al.}, \prl {\bf 88}, 257005 (2002).

\bibitem{kivelson}
S.A. Kivelson {\it et al.}, \prb {\bf 66}, 144516 (2002).

\bibitem{podolsky}
D. Podolsky {\it et al.}, \prb {\bf 67}, 094514 (2003).

\bibitem{vafek}
H.D. Chen {\it et al.}, \prl {\bf 93}, 187002 (2004).

\bibitem{congjun}
C. Wu {\it et al.}, \pra {\bf 69}, 043609 (2004).

\end{thebibliography}
\end{document}